\documentclass[preprint,showpacs,preprintnumbers,
amsmath,showkeys,amssymb,aps,floatfix]{revtex4}

\usepackage{graphicx}
\usepackage{dcolumn}
\usepackage{bm}
\usepackage{longtable}
\usepackage{epsfig}
\usepackage{times}
\usepackage[dvips]{color}
\begin{document}

\title{Coherent control of nuclear forward scattering}

\author{Adriana \surname{P\'alffy}}
\email{Palffy@mpi-hd.mpg.de}

\author{J\"org \surname{Evers}}
\email{Joerg.Evers@mpi-hd.mpg.de}

\affiliation{Max-Planck-Institut f\"ur Kernphysik, Saupfercheckweg~1, 
69117 Heidelberg, Germany}
\date{\today}

\begin{abstract}
The possibility to control the coherent decay of resonant excitations in nuclear forward scattering  is investigated. By changing abruptly the direction of the nuclear hyperfine magnetic field, the coherent scattering of photons can be manipulated and even completely suppressed via quantum interference effects between the nuclear transition currents. The efficiency of the coherent decay suppression and the dependence of the scattered light polarization on the specific switching parameters is analyzed in detail. Using a sophisticated magnetic switching sequence involving four rotations of the hyperfine magnetic field, two correlated coherent decay pulses with different polarizations can be generated out of one excitation, providing single-photon entanglement in the keV regime. The verification of the generated entanglement by testing a single-particle version of Bell's inequality in an x-ray optics experimental setup is put forward.







\end{abstract}

\maketitle

\section{Introduction}
The development of coherent light sources in the optical regime has boosted atomic physics and quantum optics in the last decades, opening new directions towards control of the atomic dynamics exploiting coherence and interference effects. 
In nuclear physics, coherent control of nuclear excitations has been a long-time goal, 
as it is related to a number of promising applications such as nuclear 
quantum optics~\cite{nqo1,nqo2,isomers1,isomers2,nuclear_laser1,nuclear_laser2,Odeurs}, including isomer depletion~\cite{isomers1,isomers2} or the  problem of gamma ray lasers~\cite{nuclear_laser1,nuclear_laser2,Odeurs}. However, nuclei proved to be much more difficult to control, due to the lack of coherent gamma ray sources and the more complicated underlying internal nuclear structure.  Coherent gamma-ray optics, for years limited by the  low intensity of the M\"ossbauer sources, was  rendered possible with the advent of high brightness synchrotron radiation sources. Nuclear forward scattering (NFS) of synchrotron radiation (SR)~\cite{Hastings,NFSReviews}  is a well-developed experimental setup in which coherent control has been achieved, by exploiting the properties of delocalized excitations in systems of identical particles.

Resonant scattering of SR on a nuclear ensemble,  such as identical nuclei in a crystal lattice, occurs via an intermediate excited state which is excitonic in nature \cite{Tramell_book,AK,Bible,Ralf}. The decay of this collective nuclear excited state then occurs coherently in the forward direction, giving rise to NFS, and in the case of nuclei in a crystal also at  Bragg angles  \cite{Kagan,Bible,Smirnov}. The   
correlation of nuclear excitation amplitudes in the excitonic state also leads to 
a speed up of the coherent decay, and thereby to a relative suppression of incoherent decay channels~\cite{Kagan_JPC,Smirnov,vanB_PRL59,Hastings}. The coherent decay channel thus becomes considerably faster than the spontaneous  one (characterized by the natural lifetime of a single nucleus) and  can be used as control mechanism of the resonant excitation in NFS.

Control of the coherent decay of the nuclear exciton enabled the observation of gamma echos in NFS experiments originally using M\"ossbauer sources \cite{gammaecho}, followed by the observation of nuclear exciton echos produced by ultrasound in forward scattering of SR \cite{ultrasound}. Here the coherent decay was  manipulated via the relative phase between the electromagnetic field scattered from two sample foils. Alternatively, changing the hyperfine magnetic field at the nuclear sites also provides a way to control the coherent decay. Following the experiment described in Ref.~\cite{JPhysCondM} on the effect of an abrupt reversal of the hyperfine magnetic field direction for NFS of light from a M\"ossbauer source, 
 results confirming the feasibility of nuclear coherent control also in NFS of SR were presented in Ref. \cite{Shvydko_MS}. The decay rate of $^{57}\mathrm{Fe}$ nuclei in a $^{57}\mathrm{FeBO}_3$ crystal excited by 14.4~keV SR pulses was changed by switching the direction of the crystal magnetization. 
The nuclear hyperfine fields were used to partially switch the coherent decay channel of the nuclear excitation off and subsequently on, demonstrating the possibility to store nuclear excitation energy. The partial suppression and subsequent release of the coherent nuclear decay are the consequence of interference between the hyperfine transitions, bearing a close resemblance to the underlying effect of electromagnetic induced transparency in quantum optics \cite{Scully}. 
While in Ref.~\cite{Shvydko_MS} only partial suppression of the coherent decay was experimentally achieved, it was pointed out that generally complete suppression should be possible if the magnetic field is switched in a direction parallel to the incoming SR light. However, the suppression is very sensitive to the exact timing of the magnetic field rotations.

In this paper we investigate the effect of the  switching moment (from here on denoted  {\it switching time}, not to be confused with the duration of the switching process) on the coherent decay intensity and polarization. We focus in particular on the scattered light in a sequence of switchings that turn the coherent decay off and on,  and study the possibility to control  the polarization of the scattered light by selecting the appropriate switching times.  An extensive analysis of the time dependence of the individual polarization components during the magnetic switching sequence allows us to determine the parameters for which two time-resolved correlated coherent decay pulses with different photon polarizations are obtained out of one incident SR pulse.

As it has been pointed out in Ref.~\cite{ourPRL}, in conjunction with the fact that the prompt SR pulse typically creates only one (and more often no) resonant excitation in the target \cite{PotzelPRA63}, the design of this sequence of magnetic switchings provides a coherent control scheme to generate keV single-photon entanglement in a NFS setup. The excitonic state produced by the SR pulse, consisting of one delocalized excited nucleus, can decay (apart from losses) in either of the two 
time-resolved correlated coherent decay pulses with different photon polarizations. The single photon corresponding to the nuclear decay entangles the two differently polarized field modes. This is single-photon entanglement, often also called entanglement with vacuum, that does not involve entanglement of two particles, but of two spatially or temporally separated field modes. Since here the field modes, and not the associated photons, play the role of the information and entanglement carriers, i.e., qubits (called in this case ebits), there is only one photon involved. This peculiarity in view of the traditional entanglement of two particles  generated some debate \cite{spe1,spe2,spe3,spe4}.

Due to the possible applications, and not least in view of the controversy related to single-photon entanglement, tests to confirm the generation of  keV-photon entanglement from a NFS setup are of great relevance. Usually, proof of entanglement is provided by the violation of a Bell-type inequality (which in this case should be however adapted for single-particle entanglement). Alternatively, applications such as successful teleportation experiments can indicate entanglement~\cite{Lee}. Here we propose and discuss an experimental setup using modern x-ray optics devices such as  polarization-sensitive mirrors, phase shifters and x-ray beam splitters, that can test the single-particle version of Bell's inequality. An experimental realization and verification of keV single-photon entanglement would bring quantum information in a new energy regime, and could open new perspectives for the creation  and verification of quantum superposition in  micro-optomechanical systems \cite{Vahala,Marquardt}, or alternatively robust quantum key distribution \cite{LeeArxiv} and quantum cryptography \cite{LeePRA} using entangled field modes of a single x-ray photon.

\section{Magnetic switching in NFS}

Let us consider SR light scattering  on nuclei inside a crystal target in a  magnetic field  that provides  hyperfine splitting of the nuclear levels according to their spin. The energy of the SR pulse is  chosen  such that it can resonantly drive  a nuclear transition from the nuclear ground state to the first excited state. 
Because both the  duration and the transit time of the SR pulse shining on a  crystal target are short
compared to the excited state lifetime $\tau$  of the nuclear excited state, the pulse creates a collective nuclear excited state  which is a spatially coherent superposition of the various excited state hyperfine levels of all the nuclei in the crystal. The SR pulse is considered in the following to be a delta peak.

In an excited state, an isolated nucleus will decay via radiative decay 
or  by internal conversion decay (IC), with the total decay width  $\Gamma=\Gamma_{\gamma}+\Gamma_{IC}$.  For the 14.4 keV first excited level of $^{57}\mathrm{Fe}$, the traditional M\"ossbauer excitation used in most NFS experiments, $\Gamma_{IC}/\Gamma_{\gamma}\simeq 8$, so that an isolated excited nucleus is eight times more likely to give up its energy via IC. However, for a collection of nuclei, due to the formation of the exciton, the probability for radiative decay can be greatly enhanced by spatial coherence \cite{Bible}. The incident wave is scattered in the forward direction (we will assume in the following only NFS, and no Bragg scattering), and speeded-up compared to the incoherent decay. 

\begin{figure}
\begin{center}
\includegraphics[width=0.7\textwidth]{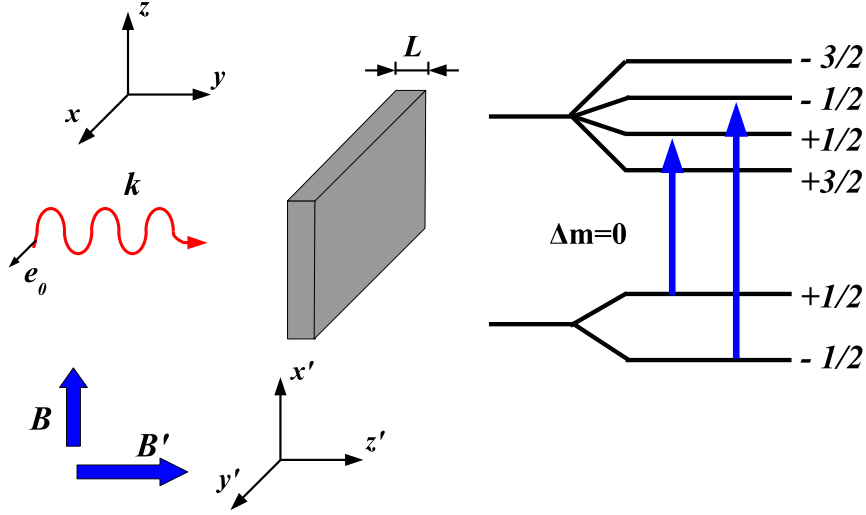}
\caption{\label{geometry} Scattering geometry  and hyperfine level scheme of the $^{57}\mathrm{Fe}$ transition. The incident radiation has the wave vector $\vec{k}$ parallel to the $y$ axis and the polarization $\vec{e}_0$ parallel to the $x$ axis. The initial magnetic field $\vec{B}$ is oriented parallel to the $z$ axis. The switching that can completely suppress the first order scattering is achieved by a rotation that  brings the crystal magnetization and the hyperfine field $\vec{B}'$  in the  direction of the incoming radiation. The new reference frame is then $\{x', y', z'\}$.}
\end{center}
\end{figure}

In the presence of a strong  magnetic field, the $^{57}\mathrm{Fe}$ nuclear ground and excited state (of spins $I_g$=1/2 and $I_e$=3/2, respectively) split into hyperfine levels, as shown schematically in  Figure \ref{geometry}. Depending on the geometry of the setup (see Figure \ref{geometry}) and the polarization of the incident SR radiation, some of the six possible hyperfine transitions will be excited, with $\Delta m=m_e-m_g=\pm 1$ or $\Delta m=0$, where $m_g$ and $m_e$ are the projection of the nuclear spin of the ground and excited states, respectively, on the quantization axis. The coherent scattering implies that each nucleus decays back to its original  hyperfine level $m_g$, such that the initial and the final states of the nuclei coincide. 

There are several theoretical approaches to treat the coherent nuclear excitations induced by SR pulses and calculate the amplitude of the scattered light. Because of the different time scales, the formation of the intermediate excitonic state and its subsequent decay can be treated as two independent quantum mechanical processes, or, alternatively, the whole interaction can be considered as a scattering problem \cite{Tramell_book,AK}.  
In the following we briefly present a simple semiclassical approach based on finding the solution of the scattering problem directly in time and space \cite{Shvydko_theory}. This method is chosen because it allows for the implementation of the switching in the calculation in a straight-forward way.

 The amplitude  $\vec{E}(y,t)$ of the radiation pulse caused by the coherent forward scattering from the $^{57}\mathrm{Fe}$ nuclei in the sample can be calculated using the wave equation \cite{JPhysCondM,Shvydko_theory}
\begin{equation}
\frac{\partial\vec{E}(y,t)}{\partial y}=-\sum_{\ell} K_{\ell} \vec{J}_{\ell}(t)\int_{-\infty}^t d\tau \vec{J}^{\, \dagger}_{\ell}(\tau)\cdot \vec{E}(y,\tau)\, .
\label{wave_eq}
\end{equation}
Here, $\ell$ is a summation index running over $m_e$ and $m_g$ and the nuclear site.  We assume for simplicity in our calculation only one nuclear site (for a more quantitative calculation taking into account particular nuclear sites or sample characteristics, see the MOTIF code \cite{Motif}).  The excitation and decay steps of the resonant scattering are represented by the nuclear transition current matrix elements $\vec{J}_{\ell}(t)$. For a constant hyperfine magnetic field $\vec{B}_0$, the nuclear current matrix element can be written as \cite{Shvydko_MS}
\begin{equation}
 \vec{J}_{\ell}(t)=\vec{j}_{\ell}(\vec{k})\,e^{-i\Omega_{\ell}t-\Gamma_0t/(2\hbar)}\, ,
\label{current}
\end{equation}
where  $\hbar\Omega_{\ell}$ is the correction to the transition energy $E_0=\hbar\omega_0$ due to the hyperfine interaction and $\Gamma_0$ is the natural width of the nuclear excited state. The hyperfine correction frequencies  $\Omega_{\ell}$ were obtained from the MOTIF code \cite{Motif,Shvydko_theory} for 300~K temperature omitting the weak polarization mixing. The constant  in Eq.~(\ref{wave_eq}) is defined as $K_{\ell}=2\pi N_n f_n(\vec{k})/[kc^2(2I_g+1)]$, with $N_n$ the number of nuclear sites per unit volume, $k$ the wave number and $f_n(\vec{k})$ the M\"ossbauer factor. 
The nuclear current density matrix elements in the momentum representation $\vec{j}_{\ell}(\vec{k})$ for $M1$ transitions are given by \cite{JPhysCondM,Shvydko_theory,YuriHInt}
\begin{eqnarray}
\vec{j}_{\ell}(\vec{k})&=&\langle I_g m_g|\vec{j}(\vec{k})|I_em_e\rangle
\nonumber \\
&=&\left[\frac{3(2I_e+1)c^5 \Gamma_{\gamma}}{4\omega_0^3}\right]^{1/2}\sum_{q=0,\pm 1} \left(\begin{array}{c c c} I_g\   & 1& \ I_e \\ -m_g \ & q & \ m_e \end{array} \right)(-1)^q \vec{k}\times \vec{n}_{-q}\, ,
\label{currents}
\end{eqnarray}
with the spherical unit vectors $\vec{n}_q$ defined in terms of the cartesian unit vectors
\begin{subequations}
\begin{align}
\vec{n}_0&=\vec{e}_z\, ,  \\
\vec{n}_{+1}&=-\frac{1}{\sqrt{2}}(\vec{e}_x+i\vec{e}_y)\, , \\
\vec{n}_{-1}&=\frac{1}{\sqrt{2}}(\vec{e}_x-i\vec{e}_y)\, .
\end{align}
\end{subequations}
The nuclear current matrix elements correspond to the particular polarization component $q=0,\pm 1$ of the nuclear transition, with $q=m_e-m_g$ dictated by the properties of the 3-$J$ symbol. The scalar product $\vec{J}^{\, \dagger}_{\ell}(\tau)\cdot \vec{E}(y,\tau)$ in Eq.~(\ref{wave_eq}) reduces the sum $\sum_{\ell}$ over the ground and excited state hyperfine levels to those $\ell$ indices (and therefore $m_g$ and $m_e$ quantum numbers) that correspond to the initial SR pulse polarization. In our example, the initial pulse polarization is along the $x$ axis (as shown in Figure~\ref{geometry}) and can only  excite $\Delta m=0$ nuclear transitions.   

\begin{figure}
\begin{center}
\includegraphics[width=0.7\textwidth]{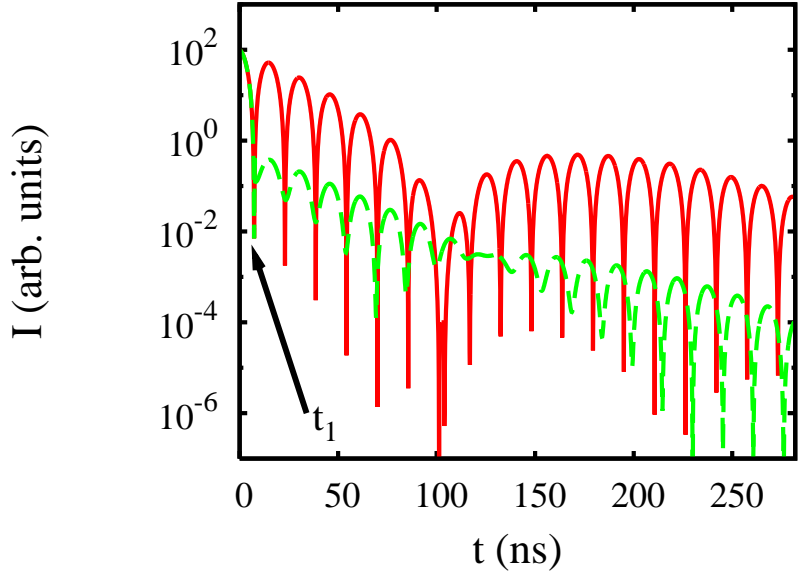}
\caption{\label{intensity} Calculated temporal evolution for the coherent nuclear decay:  unperturbed (red line) and suppressed (dashed green line) by a rotation of the hyperfine magnetic field to a direction perpendicular to the sample at time $t_1$=8~ns. The switching time corresponds to the first minimum of the quantum beat. } 
\end{center}
\end{figure}

Equation (\ref{wave_eq}) can be solved iteratively starting from the incident pulse of amplitude $\vec{E}_0(0,t)=\mathcal{E}\delta (0) \vec{e}_0$  to yield a sum 
\begin{equation}
\vec{E}(y,t)=\sum_{n=0}^{\infty}\vec{E}_n(y,t)\, ,
\label{E_sum}
\end{equation}
 in which each term represents a multiple scattering order. 
The total intensity of the scattered radiation due to the coherent decay channel is then given by  $I(t)\propto|\vec{E}(L,t)|^2$, where $L$ is the thickness of the sample. 
Since the incident SR term $\vec{E}_0$ only plays a role at t=0, we neglect it in the calculation of the intensity. As an example, the intensity of the coherently scattered SR radiation for the 14.413~keV M\"ossbauer transition in $^{57}\mathrm{Fe}$ for a sample of effective thickness $\xi=\sigma_R N_0 L/4$=5 is presented in Figure~\ref{intensity}. The effective thickness is determined by the radiative nuclear resonance cross-section $\sigma_R$, the number density of $^{57}\mathrm{Fe}$ nuclei in the sample $N_0$ and the sample thickness $L$. As  expected, in Figure~\ref{intensity}  two beat patterns can be observed: the quantum beat due to the interference between the two transitions driven by the SR pulse, given by the sum over $\ell$ in Eq.~(\ref{wave_eq}), and the dynamical beat due to multiple scattering terms, given by the sum over the scattering orders $n$  in Eq.~(\ref{E_sum}). Note that the excitation losses via the incoherent radiative and IC decay channels are included in the calculation by the natural exponential decay term in the expression of the nuclear currents (\ref{current}).

Magnetic switching denotes an abrupt rotation of the hyperfine magnetic field   direction that changes the hyperfine interaction Hamiltonian and the nuclear sublevels basis.  
Subsequently, a new quantization  axis $z'$ and therefore new eigenvectors for the hyperfine Hamiltonian are formed.
Since only the direction and not the magnitude of the magnetic field is changed, the magnetic switching corresponds 
 to a redistribution of the nuclear state populations between the hyperfine levels. Each monochromatic transition $\ell$ between the original hyperfine levels is then transformed into a multiplet composed of all allowed transitions $\ell'$ projected onto the new quantum axis $z'$.
The transformation  of the nuclear current matrix elements $\vec{J}_{\ell}(t)$ is related to the transformation of the  basis of nuclear hyperfine levels $|I m\rangle$ with the rotation of the quantization axis. According to the  representation of finite rotations \cite{Edmonds}, the rotation  $D(0 \, \beta \, 0)$ (with the Euler angles $\alpha=0$, $\beta$ and $\gamma=0$) of the quantization axis transforms the  spin eigenvectors  via the  rotation matrix elements $d^{I}_{mm'}(\beta)=\langle I\, m|D(0 \, \beta \, 0)|I\, m' \rangle$,
\begin{equation}
|I\, m'\rangle =\sum_{m} d^{I}_{mm'}(\beta)|I\, m \rangle\, .
\end{equation}
Here $I$ is the nuclear spin and $m$ and $m'$ its projections  on the old and new quantization axes $z$ and $z'$, respectively, and $\beta$ the magnetic field rotation angle. If the magnetic switching has occurred at $t=t_0$, the  nuclear current after the switching can be written as \cite{YuriHInt,Shvydko_MS}:
\begin{equation}
\label{xyz}
\vec{J}_{\ell^\prime} (t)= e^{-i \Omega_{\ell^\prime} t_0} \sum_{\ell}
d^{I_e}_{m^\prime_e m_e} (\beta)  d^{I_g}_{m^\prime_g m_g} (\beta)
\vec{J}_\ell(t-t_0)\, .
\end{equation}
The rotation matrix elements $d^{I_e}_{m^\prime_e m_e} (\beta)$ with $m_e, m^\prime_e \in \{-3/2,-1/2,1/2,3/2\}$ form a $4\times 4$ dimensional set, while the corresponding set of $ d^{I_g}_{m^\prime_g m_g} (\beta)$ with $m_g, m^\prime_g \in \{-1/2,1/2\}$ has dimension $2\times 2$. Thus, each new nuclear current is  a linear combination of all six old currents, with coefficients given by the rotation matrix elements for the ground and excited-state hyperfine levels $d^{I_e}_{m^\prime_e m_e} (\beta)  d^{I_g}_{m^\prime_g m_g}(\beta) $. Consequently, the vectorial orientation of the nuclear currents  and thus  the polarization of the scattered light change under the magnetic field rotation. Furthermore, the new currents $\vec{J}_{\ell^\prime}(t)$ corresponding to each original excitation transition can interfere, and depending on the switching time $t$ and the switching angle $\beta$, their interference can be constructive or destructive. Suppression or restoration, i.e., control of the coherent nuclear decay, can thus be achieved by optimizing the time and angle of switching. 

Experimentally, prompt magnetic switching is possible in crystals that allow for fast rotations of the strong crystal magnetization via weak external magnetic fields. The switching experiment in Ref.~\cite{Shvydko_MS} was facilitated by $^{57}\mathrm{FeBO}_3$, a canted antiferromagnet with a plane of easy magnetization parallel to the (111) surfaces. Initially, a constant weak magnetic field induces a magnetization parallel to the crystal plane surface and aligns the magnetic hyperfine field $\vec{B}$ at the nuclei. The magnetic switching is then achieved by a stronger, pulsed magnetic field in a  crystal plane perpendicular to the original magnetic field direction, that rotates the magnetization by an angle $\beta$ and realigns the hyperfine magnetic field.  Because of the perfection of the crystal, the desired rotation of the magnetization occurs abruptly, over less than 5~ns~\cite{Shvydko_EPL}.

A rotation of the magnetic field in the sample plane with an angle $\beta$ ranging up to $72^{\circ}$ has been investigated in the experiment described in Ref.~\cite{Shvydko_MS}. By changing the direction of the hyperfine magnetic field at a minimum of the unperturbed quantum beat pattern,  partial suppression of the coherent decay channel was achieved, whereas by switching at a maximum the polarization of the scattered light is changed. In this case, the original $\Delta m=0$ frequency components were transformed to $\Delta m=\pm 1$ components after the switching.

In the following, we analyze the dependence of the switching as a function of the switching time and angle considering the first order scattering. The first scattering term $n=1$ in the expansion Eq.~(\ref{E_sum}) corresponds to the thin crystal limit and gives the largest contribution to the scattered light for our parameters.  The incident radiation pulse excites the monochromatic nuclear transitions $\ell$. The  scattered electric field obtained from the wave equation (\ref{wave_eq}) can then be written as
\begin{equation}
\vec{E}_1(y,t)=-y\mathcal{E}\sum_{\ell}\vec{\mathcal{A}}_{\ell}(\vec{k})\,e^{-i\Omega_{\ell}t-\Gamma_0t/(2\hbar)}\, ,
\label{E1}
\end{equation}
with the time-independent amplitudes $\vec{\mathcal{A}}_{\ell}(\vec{k})=\vec{j}_{\ell}(\vec{k})K_{\ell}[\vec{j}^{\dagger}_{\ell}(\vec{k})\cdot \vec{e}_0]$. Each component $\ell$ corresponds to a certain polarization of the scattered light. After switching at time $t_1$ with a  rotation of  the hyperfine magnetic field in the crystal plane by the angle $\beta$,  the amplitudes are build up via the interference of the initially excited transitions $\ell$ \cite{Shvydko_MS}, 
\begin{eqnarray}
\vec{\mathcal{A}}_{\ell'}(\vec{k})&=&\vec{j}_{\ell'}(\vec{k})K_{\ell'}\sum_{\ell}
d_{m_g'm_g}^{I_g}(\beta)d_{m_e'm_e}^{I_e}(\beta) \,
[\vec{j}^{\dagger}_{\ell}(\vec{k})\cdot \vec{e}_0] \, e^{-i\Omega_{\ell}t_1}\, .
\end{eqnarray}
As it can be seen in the above equation, the amplitudes (although otherwise time-independent) are dependent on the moment when the switching of the quantization axis took place, given by $t_1$.
If one can find the specific switching angle $\beta$ and time $t_1$ for which all six amplitudes $\vec{\mathcal{A}}_{\ell}(\vec{k})$ (corresponding to $\Delta m=0$ and $\Delta m =\pm 1$) are zero, the first order scattering after the switching, given by the field in Eq.~(\ref{E1}) with amplitudes $\vec{\mathcal{A}}_{\ell'}(\vec{k})$, is completely suppressed. Such a complete suppression, i.e., simultaneous cancellation of all  $\vec{\mathcal{A}}_{\ell'}(\vec{k})$, cannot be achieved by rotating the magnetic field in the sample plane,  irrespective of the switching time and rotation angle. The reason is that one cannot find a single  switching time that yields all amplitudes $\vec{\mathcal{A}}_{\ell'}(\vec{k})$ zero. 
It can however occur when switching the magnetic field direction parallel to the incoming photon $\vec{k}$~\cite{Shvydko_MS}, equivalent to the hyperfine basis transformation described by the rotation matrix
\begin{align}
\langle I m'| D(-\pi/2\ -\pi/2\ 0)|I m\rangle=e^{-im\pi/2}\:d^I_{m'm}(-\pi/2)\,.
\end{align}
This switching is advantageous since the new direction of the quantization axis is parallel to the light propagation direction and the $\Delta m'=0$ transitions ( which correspond to light polarization parallel to the quantization axis) are therefore not allowed. 
The new transition amplitudes after this  switching  are given by
\begin{eqnarray}
\vec{\mathcal{A}}_{\ell'}(\vec{k})&=&\vec{j}_{\ell'}(\vec{k})K_{\ell'}\sum_{\ell}
d_{m_g'm_g}^{I_g}(-\pi/2)d_{m_e'm_e}^{I_e}(-\pi/2)
\nonumber \\
&\times&
[\vec{j}^{\dagger}_{\ell}(\vec{k})\cdot \vec{e}_0]
\,e^{-i\Omega_{\ell}t_1-i(m_g'+m_e')\pi/2} \, .
\end{eqnarray}
By imposing that the remaining four new amplitudes $\vec{\mathcal{A}}_{\ell'}$ after the switching are simultaneously zero, we obtain for the  switching time $t_1$ the expression $t_1=(n-1/2)\pi/\Omega_0$, where $n\in\{1,2,\ldots\}$ and $\hbar\Omega_0$ is the hyperfine energy correction for the $\Delta m=0$ transitions. This corresponds to switching at the minima of the unperturbed quantum beat pattern, as shown by the dashed green line in Figure \ref{intensity}, calculated with a suppressing switching at $t_1=\pi/(2\Omega_0)$ and a sample of effective thickness $\xi=5$. Since all the new transition amplitudes are zero, a significant  suppression of the coherent decay channel is achieved, so that excitation energy is stored inside the crystal. The intensity is not exactly zero due to the small non-zero contributions of the higher scattering order terms in Eq.~(\ref{E_sum}). In principle, the same procedure can be applied to find the proper switching time which suppresses completely the second order scattering (i.e., $\vec{E}_2$ in Eq.~(\ref{E_sum})), third order scattering, etc. However, the optimal switching times differ from one scattering order to another, so that only one scattering order can be suppressed at a time. In the following 
we restrict our treatment to suppression of the dominant first order scattering.

\begin{figure}
\begin{center}
\includegraphics[width=0.7\textwidth]{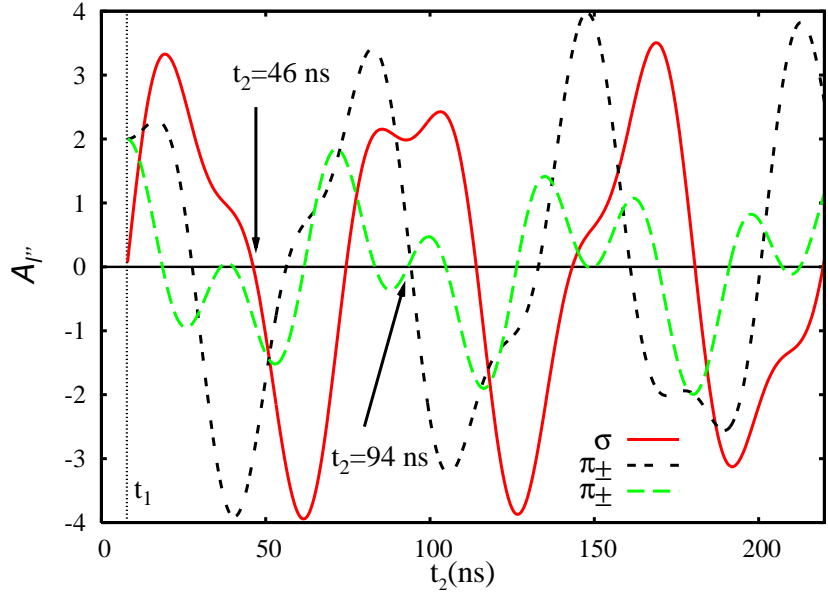}
\caption{\label{amplitudes} Transition amplitudes after 2 switchings as described in the text at times $t_1=\pi/(2\Omega_0)$ and $t_2$, as a function of $t_2$. Red full line presents up to a factor the $(1/2\!\to\! 1/2)$ and $(-1/2\!\to\! -1/2)$ transition amplitudes, black double dashed line, the $(-1/2\!\to\! 1/2)$ and $(1/2\!\to\! -1/2)$ transitions and the green dashed line the  $(3/2\!\to\! 1/2)$ and $(-3/2\!\to\! -1/2)$ transitions, respectively.}
\end{center}
\end{figure}

The various frequency components can be recovered by switching back the magnetic field to the initial direction parallel to the original $z$ axis at a proper instant in time $t_2$. The new amplitudes after this second switching depend on the two switching times  $t_1$ and $t_2$. 
Assuming that the first switching occurred at $t_1=\pi/(2\Omega_0)$, the dependence of the six new amplitudes $\vec{\mathcal{A}}_{\ell''}(\vec{k})$ on the second switching $t_2$ determines the polarization of the scattered light.  
In  Figure~\ref{amplitudes} we show  the amplitudes $\vec{\mathcal{A}}_{\ell''}(\vec{k})$ after a suppressing switching at time $t_1=\pi/(2\Omega_0)$ and a reverse switching  at time $t_2$ as a function of when the second switching occurred. Out of the six non-zero components, the two components $\vec{\mathcal{A}}_{\ell''}(\vec{k})$ corresponding to $\Delta m = 0$  transitions $(-1/2\! \to \! -1/2)$ and $(1/2\! \to \! 1/2)$ are proportional to $3\sin \Omega_1(t_2-t_1)+\sin \Omega_0(t_2-t_1)$, while the $\Delta m=\pm 1$ amplitudes  for transitions $(-1/2 \! \to \! 1/2)$ and $(1/2 \!\to\! -1/2)$ are proportional to $3\cos\Omega_1(t_2-t_1)-\cos\Omega_0(t_2-t_1)$ and the ones for $(3/2\!\to\!1/2)$ and $(-3/2\!\to\!-1/2)$ to $\cos\Omega_1(t_2-t_1)+\cos\Omega_0(t_2-t_1)$, respectively.  Here  $\hbar\Omega_1$ denotes the hyperfine energy correction for the  $(-1/2\!\to\! 3/2)$ transitions.
Choosing the switching time  $t_2=46$~ns (when the amplitudes of the $\Delta m=0$ components are zero) sets free only the $\Delta m=\pm 1$ frequency components, whereas the ones with $\Delta m=0$ are still suppressed. Alternatively, for $t_2=94$~ns, for which the four amplitudes $\vec{\mathcal{A}}_{\ell''}(\vec{k})$ of the $\Delta m=\pm 1$ transitions are zero, only 
the $\Delta m=0$ component is released. Thus, the polarization of the emitted light in the coherent decay can be controlled by choosing the proper switching time. 

The choice of the switching times is particularly sensitive when a complicated switching sequence with more than one complete suppression of the coherent decay is envisaged. This is related to the fact that the field amplitudes recall all the switching history, since all previous switching times enter their expressions. A direct consequence is that the more switchings already occurred, the more limited is the choice of following switching times for which a clear suppression of one or all polarization components can be achieved. For instance, originally complete suppression of all polarization components is guaranteed at all switching times  $t_1=(n-1/2)\pi/\Omega_0$. However, after having switched at $t_1=\pi/(2\Omega_0)=8$ ns and $t_2=46$~ns, there is only one reasonable time instance $t_3$ in the first 300~ns after the SR pulse, at approx. $t_3$=99~ns, when by rotating the magnetic field to a direction perpendicular to the sample, complete suppression is achieved. 

A 4-switching scheme  that controls the polarization of the scattered light as  proposed in Ref.~\cite{ourPRL} requires in this respect a unique choice of switching parameters. By starting with the suppressing switching at the earliest time possible (the first minimum of the quantum beat),  $t_1$=8~ns, a second switching at $t_2=46$~ns bringing the magnetic field back to its original direction parallel to the crystal sample restores only the coherent decay of the $\Delta m=\pm 1$ components, as described in the example above. Assuming the incident SR pulse to be $\sigma$-polarized (in NFS, the $\sigma$ and $\pi$ polarizations are defined only by convention, with the standard notation according to which a $\sigma$ polarized photon has the electric field vector in the horizontal plane),  after the second switching only $\pi$-polarized photons are emitted. A third rotation of the magnetic field to the direction of the incident radiation at $t_3=99$~ns suppresses again almost completely the coherent decay.  The switching time $t_3$ is chosen such that after the switching, all new currents interfere again destructively. The choice of the fourth switching time $t_4$, after which the stored excitation is released by a rotation of the magnetic field back to its original direction along the $z$ axis, determines the polarization of the emitted light. There are only two favorable times $t_4$ in the first 300~ns after the excitation that release light with a single polarization component.  

\begin{figure}
\begin{center}
\includegraphics[width=0.7\textwidth]{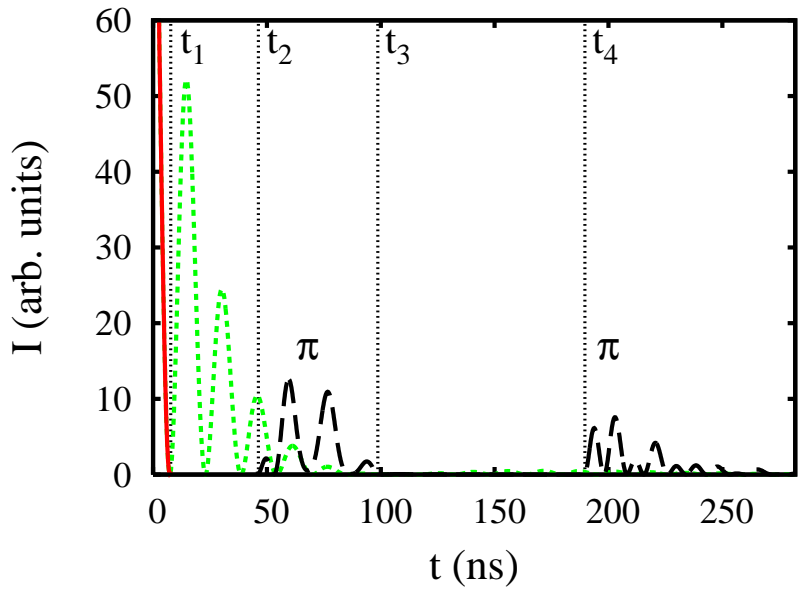}
\caption{\label{pulses2pi} Splitting of a single nuclear SR excitation into three temporally separated pulses  achieved by the magnetic switching scheme described in the text with $t_4=190$~ns. The excitation pulse and the scattered light before the switchings are $\sigma$-polarized (red solid line), while the two pulses emitted after $t_2$ and $t_4$ are $\pi$-polarized (black dashed line). For comparison, the unperturbed intensity without any switching is shown by the green dotted line. The results are calculated for an effective sample thickness of $\xi$=5.
  } 
\end{center}
\end{figure} 

For $t_4=190$~ns, all the $\Delta m=0$ amplitudes destructively interfere and a second $\pi$-polarized pulse is emitted, as shown in Figure~\ref{pulses2pi}. For comparison, we also present  the unperturbed intensity without any switching. Both spectra are calculated for an effective sample thickness of $\xi$=5.
Alternatively, following the same first three switchings at $t_1$, $t_2$ and $t_3$, a fourth switching at  $t_4=137$~ns ensures that all the $\Delta m=\pm 1$ amplitudes cancel and the emitted light is $\sigma$-polarized.  This sequence of four rotations of the hyperfine  magnetic field is therefore  designed to control the coherent decay and generate  two correlated coherent decay pulses with different photon polarizations out of one SR nuclear excitation, as first presented in Ref.~\cite{ourPRL,note}. Two  pulses  of similar intensities, the first one $\pi$- and the second one $\sigma$-polarized, are emitted  after $t_2$ and $t_4$, respectively, as shown in Figure~\ref{pulsespisigma}.   As it was argued in Ref.~\cite{ourPRL}, apart from the unavoidable coherent decay before the first switching for $0\le t \le t_1$, the two $\sigma$ and $\pi$ pulses are two entangled modes of a single photon originating from the decay of the nuclear exciton.

\section{KeV single-photon entanglement tests}

\begin{figure}
\begin{center}
\includegraphics[width=0.7\textwidth]{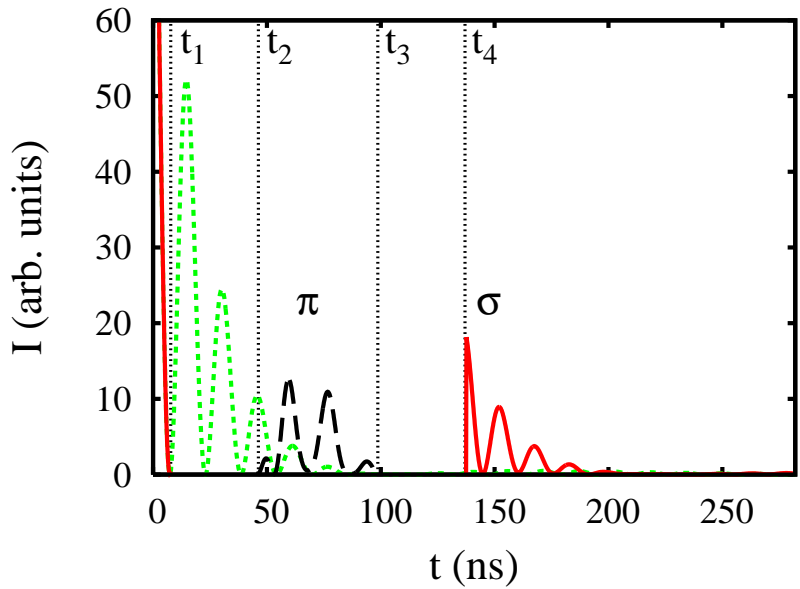}
\caption{\label{pulsespisigma} Same as Figure~\ref{pulses2pi} but with $t_4=137$~ns. Here the coherent scattering before the first switching and the second pulse are $\sigma$-polarized (red solid line), while the first pulse after $t_2$ is $\pi$-polarized (black dashed line). Since typically the SR pulse creates only a single excitation in the nuclear sample, after a background emitted at times $t\le t_1$, the two pulses of $\pi$ and $\sigma$ polarization play the role of two entangled modes of the field in single-photon entanglement. Due to the switchings, the time interval in which the coherent decay is expected is substantially shorter than the nuclear lifetime $\tau$=141~ns.  } 
\end{center}
\end{figure}

Due to the very narrow nuclear excited state width and low number of photons per mode provided by the SR source, typically one pulse creates only one or no nuclear resonant excitation in the target. Effectively, the SR thus can be compared to a weak coherent driving field in atomic quantum optics. Focussing on the single-excitation part, the nuclear exciton can be envisaged as the entanglement of $N$ nuclei by a single non-localized excitation. This single excitation will decay coherently in the forward direction or incoherently via IC or an isotropically emitted photon. 
In the case of coherent decay, the described sequence of magnetic switchings provides a splitting of the scattering response in three temporally separated peaks in  the time intervals $0\le t\le t_1$, $t_2\le t\le t_3$ and $t\ge t_4$, as shown in Figure~\ref{pulsespisigma}.
Taking into consideration only coherent decay of the single excitation at times $t\ge t_1$, the photon is either emitted with $\pi$ polarization around time $t_2$, or with $\sigma$ polarization around time $t_4$, generating  a single-photon entangled state
\begin{equation}
\label{spe}
|\Psi\rangle = \alpha|1\rangle_{\sigma}|0\rangle_{\pi} +\beta |0\rangle_{\sigma}|1\rangle_{\pi}\, .
\end{equation}
Single-particle entanglement has been discussed in 1991 by Tan {\it et al.} \cite{Tan}, who proposed that it would be possible to show a contradiction between local realism and quantum mechanics using only a single particle. The argument started a large debate in the community on whether this is entanglement at all \cite{spe1,spe2,spe3,spe4}, how one could measure particle-like or wave-like properties of the single particle at two different locations, and the need of a reference oscillator to probe single-particle entanglement \cite{Zeilinger}. One reason for this discussion  is that entanglement is commonly associated with the presence of two or more particles, whereas in single-photon entanglement the role of the particles is taken over by (potentially empty) field modes. In a certain sense, single-particle entanglement is related to  regarding the nuclear exciton as an entangled state, with one excitation equally distributed over $N$ nuclei. The difference when comparing to one photon equally distributed over $N$ field modes, however, is that the non-excited nuclei are still real particles, whereas the non-occupied field mode is part of the empty vacuum. 

Meanwhile, single-photon entanglement is recognized as the entanglement of (two) distinguishable field modes via a single photon. The two field modes typically are either spatially or temporally separated. The simplest example of single-photon entanglement is a single photon impinging on a beam splitter whose outputs are two spatially separated modes $A$ and $B$. The emerging photon is then described  exactly by the state given in Eq.~(\ref{spe}), with $A$ and $B$ replacing $\sigma$ and  $\pi$. Using spatially separated entangled field modes, successful high-fidelity teleportation experiments have been performed \cite{Lombardi,Babichev_EPL}, and homodyne tomography was used to check the nonlocality of two  optical modes entangled by a single photon \cite{Babichev}. Similarly, temporally separated field modes entangled by the presence or absence of a single photon have been demonstrated experimentally \cite{Zavatta}, and time-domain balanced homodyne detection was  performed on two well-separated temporal modes sharing a single photon \cite{Dangelo}.

In the case of the described NFS setup, the two field modes are distinguishable due to their time structure and also to their different polarizations. The different polarization of the two pulses can be used to obtain spatial separation of the entangled field modes and also the separation of the two pulses from the SR background~\cite{ourPRL}. The proposed setup makes use of  state-of-the-art x-ray polarizers and  piezoelectric fast steering mirrors. The x-ray polarizer to separate the $\pi$-polarized photon from the forward scattering setup can be provided by the Si(8~4~0) Bragg reflection with $\Theta_B=45.10^{\circ}$ which corresponds to the x-ray energy of the $^{57}\mathrm{Fe}$ first excited state. At this angle, the ratio of the integrated $\pi$ reflectivity to the integrated $\sigma$ reflectivity for a channel-cut crystal is  $10^{8}$~\cite{Toellner}. For the separation of the entangled $\sigma$ photon from the background (also $\sigma$-polarized), a piezoelectric fast steering mirror using Bragg reflections designed for specific reflection energies can be used. The temporal separation of the background (emitted at or shortly after $t=0$) and the final $\sigma$ photon emitted after $t_4$=137~ns can be exploited. A piezoelectric fast steering device can move or rotate during this time a silicon, sapphire or diamond \cite{diamonds} x-ray mirror for the 14.413 keV resonant energy of $^{57}\mathrm{Fe}$  in and out of the reflecting position.  In contrast to typical choppers, such piezoelectric switching devices, either mechanical or based on  the lattice deformation of the mirror, have the additional advantage of versatile synchronization with the incident SR.

A similar splitting of a single photon from the SR pulse via the excitonic state into two polarization modes could also be achieved otherwise, e.g.,  via scattering in the Faraday geometry~\cite{NFSReviews}, for which both $\sigma\rightarrow\pi$ and the $\sigma\rightarrow\sigma$ scattering channels are open. Such a setup is reminiscent of the standard implementation in the optical frequency regime where beam splitters are used in a similar fashion to generate the single photon entanglement. Alternatively, x-ray beam splitters can be used exactly as in the optical case. The disadvantage of such  setups for entanglement generation comes from the fact that  the photon is emitted in a rather long wave packet that has a duration on the order of the nuclear decay time $\tau$. Furthermore, in the Faraday geometry, the polarization rotation gives rise to a complicated temporal structure of the two polarization components. In contrast, the proposed coherent switching scheme emits the photon in a pulse which is not limited by the nuclear decay time, but mainly by the switching time for the magnetic field in the crystal. The photon is emitted immediately after the switchings, in time windows which are substantially shorter than the nuclear decay time. This is a crucial advantage for possible applications, for which it is desirable to have short pulses emitted at known or even controllable instances in time. This facilitates high repetition rates, and easier synchronization of different information carriers. Short pulses at known times are also important, as applications or even the verification of entanglement are typically based on coincidence or correlation measurements which are difficult if the emission time is unknown.

\begin{figure}
\begin{center}
\includegraphics[width=10cm]{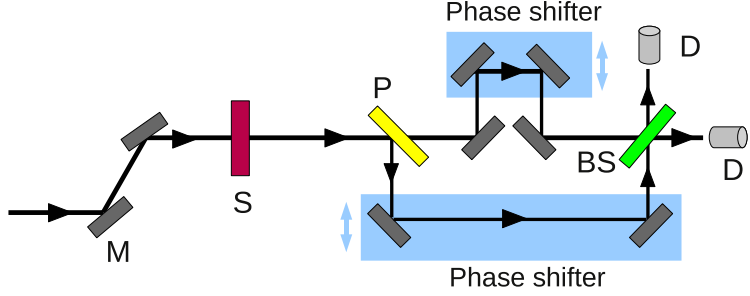}
\caption{\label{det}Entanglement detection via the violation of a Bell inequality
as proposed in Ref.~\cite{Lee}.
After the target crystal S, a polarization-sensitive  mirror P is used to
separate the two entangled field modes. The two modes are sent through
variable delay lines acting as phase shifters. Finally, the two modes
are recombined on a beam splitter BS. Photon counters D register the signal
at both output ports of the beam splitter.}
\end{center}
\end{figure}

We now turn to the question how single-photon entanglement can be verified, and how such a test can be implemented for the proposed NFS setup. An experimental test of nonlocality usually consists in performing correlation measurements on the separated field modes sharing the single photon, followed by a subsequent check if their mutual correlations violate a Bell-type inequality. A possibility to verify entanglement is offered by a single-particle version of Bell's inequality applied to an interference pattern produced by single particles~\cite{Lee}. This is a straightforward method to verify single-photon mode entanglement that  involves the non-local behaviour in interference measurements with a Mach-Zehnder interferometer. A possible adaptation to our NFS setup is shown in Figure~\ref{det}. Essentially, the coherently controlled forward scattering together with the polarization-sensitive mirror P form a beam splitter which separates a single excitation into two entangled modes. A $\pi$-polarized photon emitted at $t=t_2$ will be reflected by the polarizer P, while a $\sigma$-polarized photon emitted shortly after $t=t_4$ will be transmitted. The two modes should be recombined on a second beam splitter $BS$, forming a Mach-Zehnder interferometer.  Depending on the chosen reflections in the setup, the interferometer does not necessarily have to be a rectangular one, as it was pictured in the schematic plot in Figure~\ref{det}.  Following the recombination of the two field modes, photon counters  $D$ measure the signal from both output ports of the second beam  splitter. Since the photon is emitted in one of the two modes at different times (either shortly after  $t=t_2$ or shortly after $t=t_4$), phase shifters have to be implemented in the two arms of the interferometer. Also, in the presented setup, the initial background pulse for $t<t_1$ is not spatially separated from the signal, but rather has to be removed by time gating the detectors as in conventional NFS setups.

In Figure~\ref{det}, we chose variable delay lines as the ones under development at DESY \cite{hasylab}, that could also be used to control the temporal separation of the two signal photon components  for other entanglement applications. Alternatively, the delay lines can be replaced by phase shifters, e.g., made from silicon. In order to verify a Bell inequality violation, the photon count rates at the two output ports have to be recorded for different values of the phase shifters, and compared to contradictory theoretical predictions based on quantum mechanical calculations and on ``classical'' calculations~\cite{Lee}. 

From a technical point of view, it should be noted that different types of interferometers have already been suggested and realized in the relevant photon energy regime, see, e.g.~\cite{shvydko-again}. A Mach-Zehnder-type experiment operating at wavelength  $0.663$~{\AA} with crystal blocks forming the interferometer and Si wedges as phase shifters was performed at SPring-8~\cite{spring-8}. Laue and Bragg crystal reflections have  been used for x-ray Michelson interferometer setups \cite{Michelson1,Michelson2,Michelson3}. Thus, a verification of nonlocality in our NFS single-photon entanglement should be possible with present x-ray optics devices and SR sources.

Finally, one should mention that also other tests to verify single-photon entanglement have been performed in optical setups using a variety of Bell inequalities.  In Ref.~\cite{JDow}, it has been shown that the $N00N$ states (maximally path-entangled number states of the form $|\Psi\rangle\sim (|N\rangle_a|0\rangle_b+e^{i\varphi}|0\rangle_a|N\rangle_b)$)
with any finite number of photons $N$, including $N=1$, violate a Clauser-Horne Bell inequality.  Also, the Banaszek-Bell inequality applies to both two-particle and two-mode entangled systems, including spatially and temporally delocalized single photons  \cite{Dangelo} and achieves higher levels of violation compared to all Bell-type inequalities theoretically proposed for continuous variables. 
In Refs.~\cite{Babichev,Hessmo}, the experimental data on a single photon delocalized over two spatially separated optical modes is shown by means of homodyne tomography to violate a Bell inequality with a detection loophole which is however typical for experiments involving photon counting. 

What is characteristic for these experimental proofs of violation of Bell-type inequalities in the optical regime is the use of a local oscillator, a reference oscillator which is necessary to perform measurements of wave-like properties of the single photon. The need of such a local oscillator was one of the arguments originally used against single-particle entanglement, as the notion of {\it single} particle nonlocality was put in doubt \cite{Zeilinger}. In contrast to the optical frequency region, in the case of the  NFS setup  the phase of the SR photons is unknown, and the possibility of employing a local oscillator is therefore questionable due to technical reasons. 
The proposed Mach-Zender interferometry experiment  does not require such a local oscillator and is therefore the appropriate choice for the verification of single-photon entanglement in our NFS setup.

Once verified, one potential application of single-photon entanglement in the x-ray energy regime is the creation and verification of a quantum superposition in an optomechanical system~\cite{Vahala,Marquardt}. In such setups, one of the stationary mirrors of a cavity is replaced by a movable mirror, and the radiation pressure of the light circulating in the cavity can lead to entanglement between light and mechanical motion, or to the creation of a superposition of two macroscopic mechanical mirror states. The experimental realization of such a macroscopic quantum superposition is severely technically demanding, and keV photons could be of advantage due to their large energy and momentum compared to optical photons. 

Also, recently the advantages of single-photon entanglement compared to traditional polarization or phase encoding for quantum key distribution have been discussed \cite{LeeArxiv}. Single-photon entanglement requires on average less than one photon for each qubit and allows thus for low energy-expense encoding of quantum information. The practical disadvantage of quantum key distribution schemes involving single-photon entanglement is that photon losses (which are inevitable) lead not just to reduced key rates but also to errors caused by misidentification of the vacuum state due to experimental accuracy. The high energy and especially the high directionality of the photons emitted in NFS could be of advantage in this respect.

\section{Conclusions}
The coherent decay of the nuclear exciton in NFS can be controlled by switching the direction of the hyperfine magnetic field at the nuclei. The theoretical formalism used to describe the coherent scattering in NFS is transparent and allows the identification of the switching times and angles which lead to complete suppression and restoration of the coherent decay of one or even all polarization components to the leading order in the scattering expansion. An extended analysis of the scattered field amplitudes provides the optimal magnetic field rotation and switching time parameters for complicated switching sequences that aim at control of the time evolution of the scattered light intensity and polarization properties. As an example, a sequence of four magnetic switchings has been proposed that produces keV single-photon entanglement using nuclear excitation of the the traditional M\"ossbauer $^{57}\mathrm{Fe}$ 14.413~keV transition. A single-particle version of a Bell-type inequality can be applied to an interference pattern produced by the NFS single photons. Violation of this inequality would establish the nonlocal nature of the single-photon entangled state produced in the NFS setup. A number of applications for quantum cryptography, quantum key distribution or macroscopical quantum devices could profit from single-photon entanglement in the x-ray energy regime. 

\section{Acknowledgments}
We would like to thank Ralf R\"ohlsberger for helpful discussions.


\bibliographystyle{tMOP}
\bibliography{palffy}

\begin{thebibliography}{56}
\bibitem[39]{note}
  Note that in Ref.~\cite{ourPRL}, the switching time $t_4=190$~ns leads to
  $\pi$-polarization for both signal pulses. To achieve two signal pulses with
  different polarization, $t_4=137$~ns should be chosen.
\end{thebibliography}
\vspace{12pt}


\label{lastpage}

\end{document}